\documentclass[%
 reprint,
 amsmath,amssymb,
 aps,
]{revtex4-2}

\usepackage{graphicx}
\usepackage{dcolumn}
\usepackage{bm}
\usepackage{amssymb}
\usepackage{lipsum}
\usepackage{amsmath}
\usepackage{multirow}
\newcommand{\alc}{$\mathrm{Al/C}_{60}$}
\newcommand{\fullerene}{$\mathrm{C}_{60}$}
\DeclareMathOperator{\arsinh}{arsinh}

\begin{document}

\preprint{APS/123-QED}

\title{Effect of Interface on \\
Density and Elastic Moduli of $\textbf{Al/C}_{60}$ Nanocomposites}

\author{V.V.~Reshetniak}
\email{viktor.reshetnyak84@gmail.com}
\altaffiliation{Troitsk Institute for Innovation and Fusion Research, Russia, 108840, Moscow, Troitsk, st. Pushkovs, 12}
\affiliation{Vladimir State University named after Alexander and Nikolay Stoletov, Russia, 600000, Vladimir, Gorky st., 87}

\author{A.V.~Aborkin}%
\affiliation{Vladimir State University named after Alexander and Nikolay Stoletov, Russia, 600000, Vladimir, Gorky st., 87}

\author{A.V.~Filippov}%
\altaffiliation{Troitsk Institute for Innovation and Fusion Research, Russia, 108840, Moscow, Troitsk, st. Pushkovs, 12}

\date{\today}

\begin{abstract}
The paper analyzes theoretically the influence of fullerenes on the characteristics of \alc\ composites. The molecular dynamics method is used to study the dependences of density and stiffness constants on the concentration of inclusions, and to calculate the values of the bulk and shear moduli for isotropic polycrystalline nanocomposites. The analysis shows that interfacial interaction significantly affects the properties of nanocomposites. This effect can be taken into account within the framework of the theory of heterogeneous media using the interphase layer model. The properties of the interphase layer are determined by interfacial interaction and can be calculated by approximating the results of molecular dynamics calculations. Using assumptions about the simplified form of the two-particle distribution function allows the interfacial interaction energy to be calculated and the interphase layer properties to be assessed analytically. The paper compares numerical results and analytical estimates, and discusses the validity of the approximations used. The analysis performed on the example of an \alc\ composite material demonstrates the feasibility of using the analytical model of the interphase layer to estimate the effective density and elastic moduli of heterogeneous media with nano-inhomogeneities.
\end{abstract}

\maketitle


\section{Introduction}

Aluminum and its alloys are widely used as structural materials in aviation, nuclear industry, space industry, and automotive industry. An advantage of aluminum-based structural materials is their relatively high mechanical characteristics (stiffness, strength, and wear resistance) at low density. At the same time, very urgent are the tasks of additionally increasing strength, improving tribological characteristics, and expanding the temperature range of application of aluminum and aluminum alloy products. One of the promising approaches to solving the above problems is related to the creation of composite materials based on aluminum and reinforcing nanoparticles \cite{ceschini2017aluminum,pan2022metal,shinde2020tribological,malaki2019advanced}. Experimental studies have shown that when the concentration of nanoparticles is low (a few percent), the strength characteristics of composites can be dramatically enhanced: an increase in the yield strength, static strength, hardness, wear resistance and heat resistance \cite{aborkin2020effect,aborkin2023enhancing,aborkin2019effect,aborkin2023increasing,evdokimov2018nanostructured,evdokimov2014tribological}.

Fullerenes \fullerene\ are often used as reinforcing particles in nanocomposites, for which extremely high mechanical stiffness is theoretically predicted \cite{ruoff1991c60,khabibrakhmanov2020carbon,peon2014bond,woo1992bulk}. Using the model of an elastic deformable sphere, Ruoff and Ruoff \cite{ruoff1991c60} calculated the bulk modulus of a single \fullerene\ molecule. The obtained value of the bulk modulus, $k = 843$~GPa, turned out to be much higher than the bulk modulus of diamond (441~GPa). The bulk modulus of the \fullerene\ molecule was calculated with atomistic simulation methods by different authors. In papers \cite{khabibrakhmanov2020carbon,peon2014bond,woo1992bulk}, calculations using quantum chemistry methods yielded values of the bulk modulus in the range from 717 to 874~GPa.

Due to the small size of fullerenes, the issue of using continuum models and stiffness constants to describe their deformation raises questions. The isothermal bulk modulus of compressibility of an isotropic solid is found from the expression \cite{landau1986theory}:

\begin{equation}\label{eq1}
    k = -V \left(\dfrac{\partial p}{\partial V} \right)_T = V \left(\dfrac{\partial^2 F}{\partial V^2} \right)_T,
\end{equation} 

\noindent where $F$ is  the Helmholtz free energy, $p$ is the pressure, $V$ is the volume, and $T$ is temperature.

In considering an ensemble of fullerenes that do not interact with each other, it is possible to introduce the distribution function of carbon atoms in phase space, which will allow the temperature and specific free energy of the \fullerene\ molecule to be calculated. However, the calculation of the bulk modulus from Eq.~\eqref{eq1} turns out to be complicated because the volume of the fullerene needs to be determined. Since the radius of a particle can only be specified with an accuracy of atomic sizes, the uncertainty in calculating the volume of small nanoparticles consisting of tens or hundreds of atoms turns out to be very significant. When studying fullerenes, this uncertainty noticeably affects the calculated value of the bulk modulus, which can manifest itself, for example, in the dependence of the bulk modulus of the \fullerene\ molecule in solution on the solvent type \cite{amer2009compressibility}, or on the method for calculating the fullerene radius \cite{reshetnyak2022interatomic}.

To eliminate the indicated uncertainty, Khabibrakhmanov and Sorokin \cite{khabibrakhmanov2020carbon} proposed a method for calculating the bulk modulus of fullerenes, which involves using the values of stiffness and lengths of covalent C-C bonds. The use of this method implies knowledge only of the parameters of interatomic interaction, which can be calculated with high accuracy from the results of spectroscopic research or \textit{ab initio} computations. The value of the bulk modulus in \cite{khabibrakhmanov2020carbon} was 868~GPa, which is in good agreement with the data obtained by other authors.

Concentrations of the reinforcing phase in \alc\ nanocomposites are limited by the difficulty of uniformly distributing fullerenes throughout the volume of the matrix and, as a rule, do not exceed several percent. The smallness of particles and their concentrations makes it possible to use the equivalent homogeneity approximation, which assumes a possibility of using averaged parameter values to describe deformations of a heterogeneous medium \cite{christensen2012mechanics}. Averaging is performed over a volume that is small in comparison with the sample size, and nevertheless includes a sufficient number of nanoparticles. Calculation of averaged (or effective) characteristics of a medium (density, elastic moduli) presupposes knowledge of the characteristics of inclusions and the matrix. In considering \alc\ nanocomposites, there arises an issue about a possibility of using the calculated elastic moduli of fullerenes to predict the effective properties of composites. The results of work \cite{reshetniak2022effect}, devoted to an atomistic study of the compressibility of an aluminum melt containing fullerenes, indicate that it is inappropriate to directly use models of heterogeneous media to solve this problem. At the same time, the application of the interphase layer theory \cite{gurtin1998general,sharma2004size,lurie2006interphase} allows one to obtain satisfactory results. For approximately calculating the parameters of the interphase layer, we proposed an analytical model, which made it possible to obtain satisfactory results for this system \cite{reshetniak2022effect}. However, the effect of fullerenes on the effective elastic moduli of a solid-phase \alc\ system was not examined.

This work is devoted to the study of the elasticity of \alc\ composites at a temperature of 300~K. A model of shear strain of fullerene is proposed and the shear modulus of a single \fullerene\ molecule is calculated. The effective stiffness constants of the composites are calculated using the molecular dynamics method, and the bulk modulus and shear modulus are determined within the framework of the Voigt--Reuss--Hill approximation. A method for determining the volume fractions of fullerenes in \alc\ composites is proposed, and the dependence of the effective characteristics of the composite on the concentration of inhomogeneities is analyzed. The possibility of using the model of heterogeneous media for \alc\ nanocomposites is discussed.

\section{Research models and methods}

\subsection{Analytical Assessments}

When studying the elastic characteristics of a nanocomposite, it is reasonable to use various methods that involve numerical simulation and theoretical analysis of numerical experiments. While numerical simulation by the molecular dynamics method provides the greatest accuracy and reliability of calculation results, the use of simplified analytical models in planning numerical or laboratory experiments allows one to select a reasonable range of research parameters and to obtain preliminary estimates of the magnitude of the expected effect from the introduction of inhomogeneities. In analyzing the results of experiments or numerical simulations, the use of simple models makes it possible to explain the obtained data, evaluate the contribution of various mechanisms to the observed effects, and check the validity of the assumptions about the theory being developed.

Since the mass of an inhomogeneous substance is the sum of the masses of the components, $m = m_M + m_I$, the effective density of the composite is linearly related to the volume fraction of inclusions, $c$, by the expression

\begin{equation}\label{eq2}
    \rho = (1 - c) \rho_M  + c \rho_I,
\end{equation}

\noindent where $\rho$ is the density; the subscripts $I$ and $M$ in this work denote the properties of the inclusion (fullerene) and matrix (aluminum), respectively; and variables without subscripts are used to denote the effective properties of the composite.

Assuming that the radius of the fullerene is $R_I = 3.57$~\AA\ \cite{eletskiui1993fullerenes}, and its volume can be calculated using the formula $V_I = 4 \pi R_I^3 / 3$, we can estimate the mass density of a fullerene, $\rho_I = m_I / V_I \approx 5.99$~g/cm$^3$. Since the density of aluminum is lower (about 2.7 g/cm$^3$), one should expect a monotonic increase in the effective density of the composite with increasing concentration of fullerenes. However, taking into account the above remarks about the volume of fullerene, the quantities $V_I$ and $\rho_I$ should be treated as parameters of the model, and the described method for calculating the parameter $\rho_I$ is one of the possible methods, the applicability of which requires verification. Another possible way to calculate $\rho_I$ is to approximate the dependence of the composite density on the concentration obtained from numerical or laboratory experiments. Since the concentration of fullerenes in a composite, $c = V_I/V$, is determined from the volume $V_I$, in practice the mass fraction  $c_m = m_I / m$ is usually used, which is related to the concentration and mass density of fullerenes: $c_m = c \rho_I / \rho$. Expression \eqref{eq2} yields a linear dependence of the inverse effective density of the composite on the mass fraction of fullerenes:

\begin{equation}\label{eq3}
    \dfrac{\rho_M}{\rho} = 1 + c_m \left(\dfrac{\rho_M}{\rho_I} - 1 \right).
\end{equation}

Expression \eqref{eq3} can be used to calculate the parameter $\rho_I$ from molecular dynamics (MD) simulation results or experimental data, and to compare analytical estimates within the framework of the theory of a heterogeneous medium with numerical or laboratory experiments. The use of elasticity theory presupposes knowledge of the volume fraction of inclusions, which can be easily expressed \textit{via} the above-defined parameters:

\begin{equation}\label{eq4}
    c = \dfrac{c_m \rho_M}{c_m \rho_M  + \rho_I (1 - c_m)}.
\end{equation}

Note that knowledge of the $\rho(c_m)$ dependence is necessary for analyzing the dependence of the composite density in laboratory experiments (for example, to determine porosity using the hydrostatic weighing method). For materials produced by powder metallurgy, porosity is one of the most important parameters, which must be controlled when choosing powder consolidation regimes.

We will use the theory of a heterogeneous medium with a small volume fraction of inclusions to analytically simulate the elastic strain of a nanocomposite. Assuming that at a macroscale the composite material is isotropic and homogeneous, the effective elastic moduli can be calculated using the formula \cite{christensen2012mechanics}:

\begin{eqnarray}\label{eq5}
    k = k_M + \dfrac{(k_I - k_M)c}{1 + [(k_I - k_M)/(k_M + 4 \mu_M / 3)}, \nonumber \\
    \mu = \mu_M \left [  1 + \dfrac{15(1 - \nu_M)[(\mu_I/\mu_M) - 1] c}{7 - 5\nu_M + 2(4 - 5\nu_M)(\mu_I/\mu_M)}  \right ].
\end{eqnarray}

\noindent Hereafter, $k$ is the bulk modulus, $\mu$ is the shear modulus, and $\nu$ is Poisson's ratio.

For further analysis, in addition to the aluminum elastic moduli known from reference books, it is necessary to determine the values of the bulk modulus and shear modulus of fullerene \fullerene. Following Ref.~\cite{ruoff1991c60}, we use the model of an elastic spherical shell, the energy of which increases linearly with changing surface area:

\begin{equation}\label{eq6}
    W = \gamma \vert S_I - S_{I0} \vert,
\end{equation}

\noindent where $S_{I0}$ is the surface area of an undeformed sphere corresponding to a minimum energy; $S_I$ is the area of the deformed surface; and $\gamma$ is the specific energy, the value of which can be estimated from the structure and elastic modulus of graphite known from reference books:

\begin{equation}\label{eq7}
    \gamma = \dfrac{h}{s_{11} + s_{12}}.
\end{equation}

The values of the compliance constants $s_{11} = 0.0108$~GPa$^{-1}$ and $s_{12} = 0.0016$~GPa$^{‒1}$ are borrowed from \cite{blakslee1970elastic}, and the distance between the graphite planes, $h = 3.348$~\AA, is equal to half the height of the hexagonal crystal cell \cite{chichagov2001mincryst}. The bulk modulus is determined from the expression

\begin{equation}\label{eq8}
    k_I = -\gamma V_{I0} \dfrac{d^2 S_I}{d V_I^2}.
\end{equation}

After substituting the formulas for the surface area and volume of a sphere into \eqref{eq8}, the bulk modulus is written in the form

\begin{equation}\label{eq9}
	k_I = \dfrac{2 \gamma}{3 R_I}.
\end{equation}

Using a similar approach, it is easy to calculate the shear modulus of fullerene (see Appendix)

\begin{equation}\label{eq10}
	\mu_I = \dfrac{k_I}{5}.
\end{equation}

The elastic moduli of fullerene can be also calculated numerically. In this case, it is necessary to calculate the dependence of the particle energy on strain, and then approximate the calculation results. The particle energy can be calculated \textit{ab initio}, or using the empirical potential of interatomic interaction. In this work, both approaches were used: \textit{ab initio} calculations were performed in the Orca program \cite{neese2018software} using density functional theory (DFT), and the LAMMPS code \cite{plimpton1995fast} with the Tersoff potential \cite{tersoff1989modeling} were used to describe empirically interatomic interactions. In quantum mechanical calculations, the many-electron wave function was calculated in the TZVP basis \cite{weigend2005balanced}, and electron exchange and correlation were taken into account in the GGA-PBE approximation \cite{perdew1996generalized}. The results of the calculations performed in this work, as well as data from the works of other authors, are presented in Table~\ref{tab1}.

\begin{table*}
\centering
\caption{\label{tab1}Elastic moduli of fullerene \fullerene: \\
calculation by different methods and available literature data.}

\begin{ruledtabular}
    
\begin{tabular}{lcccc}
 & Analytical model, & DFT, & Tersoff potential, & Data from \\ 
 & this work & this work & this work & Refs.~\cite{ruoff1991c60, peon2014bond, khabibrakhmanov2020carbon, woo1992bulk} \\
 \hline
 $k_I$~GPa & 689 & 756 & 614 & 640~-- 870 \\
 $\mu_I$~GPa & 138 & 117 & 97 & --- \\
\end{tabular}

\end{ruledtabular}

\end{table*}

Analysis of the data presented in Table~\ref{tab1} allows us to conclude that there is no significant difference in the results obtained by different methods. Analytical calculation of the shear modulus overestimates the result within 35\% compared to both \textit{ab initio} computations and the data obtained using the Tersoff potential. The bulk moduli calculated in this work using various methods differ within 20\%; in the works of other authors, up to 30\%. The discrepancy may be due to both the models used in the atomistic calculation of the fullerene strain energy and the approximations used to construct analytical models. However, in our opinion, the existing error in the approximate calculation of the fullerene strain energy is not very important, since at low concentrations of fullerenes (usually several percent), this error cannot make a significant contribution to the effective characteristics of the composite material. We believe that the issue about the possibility of using calculated elastic moduli to simulate the strain of a nanocomposite is more significant.

Analysis of Table~\ref{tab1} indicates extremely high stiffness of fullerenes. All estimates indicate that the bulk modulus of fullerene is approximately 1.5-2 times higher than the bulk modulus of diamond \cite{frantsevich1982elastic}. The shear modulus is lower than that of diamond, but much higher than that of aluminum (26 GPa) \cite{frantsevich1982elastic}. Substituting these data into formulas \eqref{eq5} indicates a possibility of a noticeable increase in the elastic moduli of the \alc\ composite compared to the aluminum only, and the values of the effective elastic moduli should increase with increasing concentration.

\subsection{Molecular dynamics simulation}

In the atomistic calculation of the effective elastic moduli of the \alc\ composite, we used a hybrid model to describe interatomic interactions, which has become widespread in the simulation of metal‒carbon nanocomposites \cite{safina2020simulation}. It was assumed that aluminum atoms interact with each other \textit{via} the parameterized EAM potential \cite{mishin1999interatomic}, the interaction of carbon atoms is calculated using the Tersoff potential \cite{tersoff1989modeling}, and the Lennard‒Jones potential is used to describe the interaction of pairs of Al‒C atoms,

\begin{equation}\label{eq11}
	u(r) = 4 \varepsilon \left[ \left(\dfrac{\sigma}{r} \right)^{12} - \left(\dfrac{\sigma}{r} \right)^{6} \right].
\end{equation}

In \cite{reshetnyak2022interatomic} we calculated the potential parameters $\varepsilon = 5.2 \times 10^{-2}$~eV and $\sigma = 2.7$~\AA\ using \textit{ab initio} data on the interaction of fullerenes with aluminum surfaces. The potential was tested \cite{reshetnyak2022interatomic} by comparing the results of multiscale simulation of the kinetics of fullerene desorption from the aluminum surface with experimental data from \cite{hamza1994reaction}. The analysis showed that the use of the Lennard‒Jones potential with selected parameters makes it possible to reproduce with high accuracy the temperature dependence of the density of coating the substrate with molecules under conditions close to experimental ones. Note that the use of this parametrization is inappropriate for fullerenes of arbitrary radius, carbon nanotubes, graphite and graphene. A detailed discussion of interatomic interactions in these systems is presented in Refs.~\cite{reshetniak2020aluminum, reshetniak2023}.

\begin{figure}[ht]
\centering
\begin{minipage}{0.49 \linewidth}
	\includegraphics[width=1\linewidth]{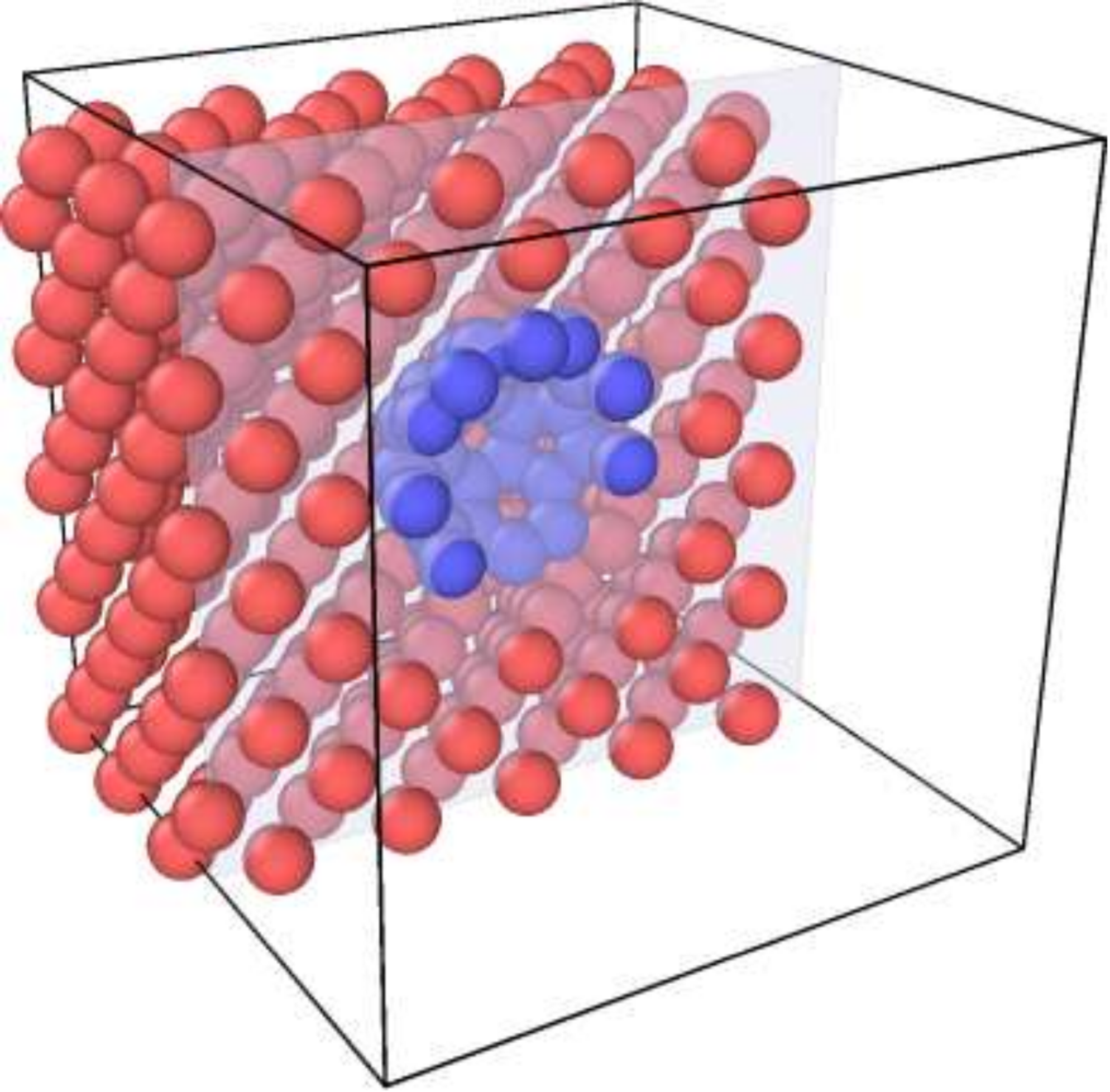}%
\end{minipage}
\begin{minipage}{0.49 \linewidth}
	\includegraphics[width=1\linewidth]{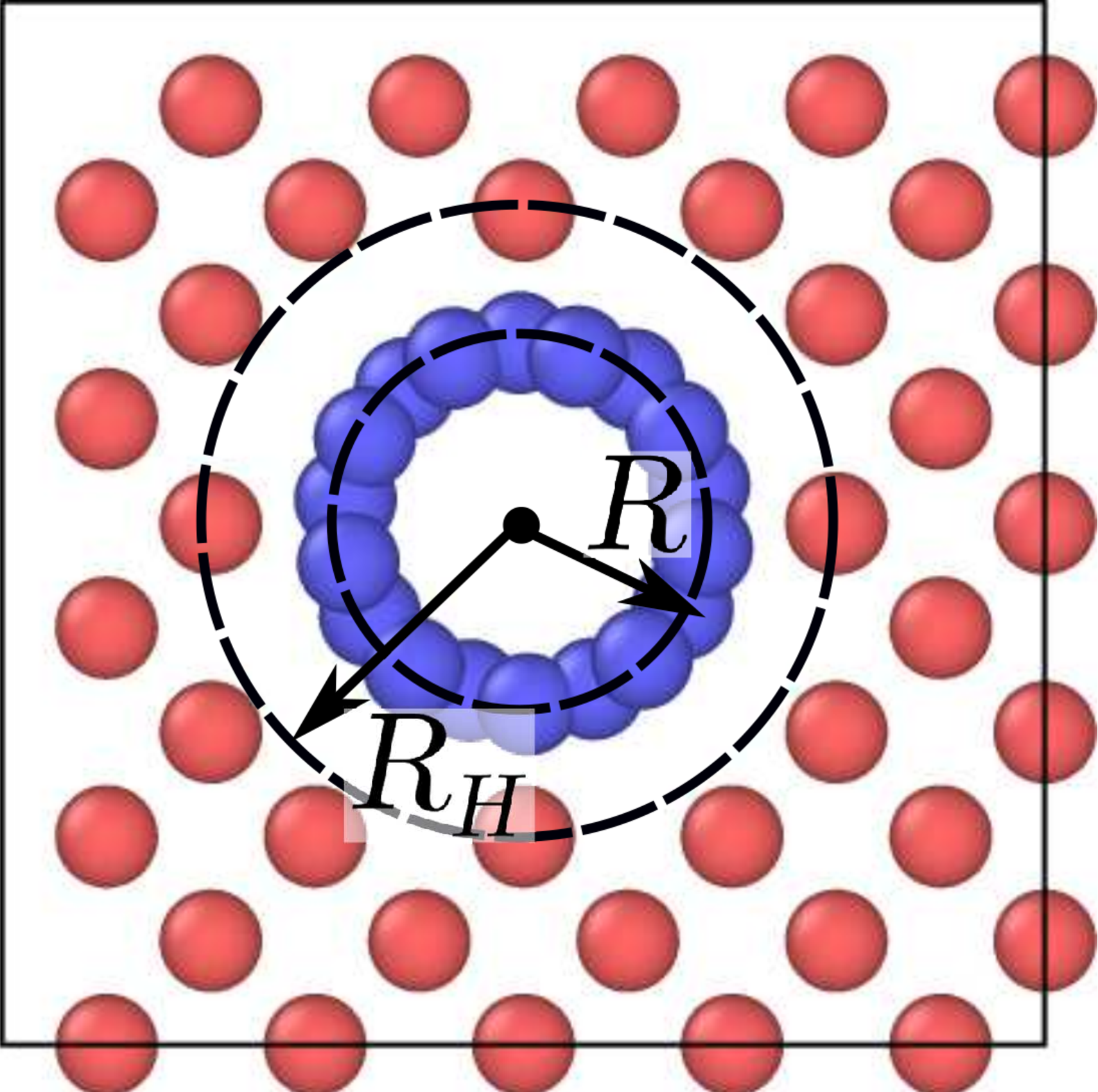}%
\end{minipage}\\
\vspace{0.2cm}

\raggedright \hspace{0.23\linewidth} $(a)$\hspace{0.45\linewidth} $(b)$

\caption{\label{fig1}
Scheme of the atomistic model of the \alc\ composite. The carbon atoms are colored in blue, and the aluminum atoms are colored in red. For clarity, only some of the atoms are shown in the figures, and the rest are removed: (a) part of the atoms is cut out by the (100) plane in a periodic cell; and (b) a thin layer of substance is cut out from the region near the center of the cell by two parallel (100) planes.
}%
\end{figure}

In the center of a cubic aluminum supercell, the size of which was calculated from a given mass fraction of carbon, a spherical cavity was cut out (Fig.~\ref{fig1}). The cavity radius is expressed as

\begin{equation}\label{eq12}
	R_H = R + (2/5)^{1/6} \sigma.
\end{equation}

Then the fullerene was placed in the cavity. Periodic boundary conditions were specified at the cell boundaries. The calculation of $R_H$ using formula \eqref{eq12} corresponds to an analytical estimate of the equilibrium value of the cavity radius at which the potential energy of the system is minimal \cite{reshetniak2022effect}. Substituting $\varepsilon = 5.2 \times 10^{-2}$~eV, $\sigma = 2.7$~\AA\ \cite{reshetnyak2022interatomic}, and $R = 3.57$~\AA\ \cite{eletskiui1993fullerenes} into formula \eqref{eq12}, we obtain $R_H = 5.88$~\AA.

Despite the fact that a cubic cell was chosen as the initial one, the relaxation of the structure and the subsequent calculation of thermodynamic quantities were carried out without any restrictions: under the influence of interatomic interaction forces, all angles and lengths of the edges of the cell could change arbitrarily in the calculations. Nevertheless, all the parameters obtained during relaxation corresponded with high accuracy to a cubic cell: the lengths of the cell edges in all cases coincided with each other with an accuracy of at least 0.017\%, and the deviation of the angles from $90^{\circ}$ did not exceed 0.008\%. Therefore, we assumed hereafter that the cell has a cubic structure.

Relaxation times in solids can significantly exceed the times available for MD simulations. Therefore, to stabilize the position of atoms at the interphase in the \alc\ composite, we used the following four-step scheme to accelerate relaxation. In the first step, to increase the mobility of atoms, the system was subjected to isostatic tension (a pressure $p = -1$~GPa) at elevated temperature (673~K). The system was stabilized under these conditions within 500~ps. Then, within the next 500~ps, the system was stabilized at a temperature of 673~K and a pressure of 2~GPa. The selected values of the state parameters are close to typical sintering conditions of Al--C composite powder materials in experiments \cite{aborkin2018effect}. Then the pressure and temperature were slowly (within 1~ns) decreased down to 0~Pa and 300~K, after which the system was stabilized within 500~ps. The average values of cell parameters, adhesion energy, density and other quantities were calculated after the system was stabilized.

It is important to note that if the model does not ensure a transition of an initially metastable system to an equilibrium state within the estimated time, the parameters of the system are determined by the prehistory, i.e., they depend on the method of stabilization and the random choice of initial velocities. In studying elastic moduli and other equilibrium properties of composites, the appearance of such effects is undesirable, and their contribution to the parameter values can be considered a model error. If the system is not sufficiently stabilized, the magnitude of this error can be very significant, exceeding the contribution of effects associated with the medium heterogeneity. The stabilization of the system can be judged by the values of the \alc\ adhesion energy, and by the linear behavior of the dependence of the effective mass density of the system, $\rho = m/V$, on the volume fraction of fullerenes, $c = V_I/V$. If relaxation is insufficient, the value of adhesion energy $U$ can vary within 10\% (about 2~eV). The use of the selected stabilization algorithm ensured the reproducibility of the $U$ value with an accuracy of 0.036\% (0.007~eV) in statistically independent calculations.

Mass density also quite reliably characterizes the stabilization of the system. If the relaxation algorithm is unsuccessfully chosen, the error in calculating the structure of the interphase affects the mass density $\rho$ more strongly than the change in the mass fraction of carbon. In this case, expression \eqref{eq3} poorly approximates the $\rho(c_m)$ dependence, which in this case turns out to be nonmonotonic. The stabilization algorithm chosen in this work provides a monotonic dependence of $\rho(c_m)$, and the accuracy of calculating the density of the system turns out to be sufficient to approximate this dependence by expression \eqref{eq3}.

After relaxation, the cell was deformed by slightly changing one of the parameters of edge length $a$ or angle $\gamma$ between the edges $a$ and $b$. The first case corresponds to specifying a nonzero component $\epsilon_{11}$ ($\epsilon_1$ in matrix notation \cite{nye1985physical}) of the strain tensor $\epsilon_{ij}$, and the second case corresponds to $\epsilon_{23}$ ($\epsilon_4$ in matrix notation). In what follows, matrix notations will be used for the strain, stress, stiffness, and compliance tensors \cite{nye1985physical}. The components of the stiffness tensor were calculated by linearly approximating the dependence of strains on stresses:

\begin{eqnarray}\label{eq13}
	\sigma_1 = c_{11}\epsilon_1, \nonumber \\
	\sigma_2 = c_{12}\epsilon_1, \nonumber \\
	\sigma_4 = c_{44}\epsilon_4.
\end{eqnarray}

Due to the cubic symmetry of the model, the elastic properties are completely determined by the three components of the compliance tensor and the calculation of the remaining components is not needed. The components of the compliance and stiffness tensors are related by the expressions \cite{nye1985physical}:

\begin{eqnarray}\label{eq14}
	s_{11} = \dfrac{c_{11} + c_{12}}{(c_{11} - c_{12})(c_{11} + 2 c_{12})}, \nonumber \\
	s_{12} = \dfrac{-c_{12}}{(c_{11} - c_{12})(c_{11} + 2 c_{12})}, \nonumber \\
	s_{44} = \dfrac{1}{c_{44}}.
\end{eqnarray}

The values of the stiffness constants obtained as functions of the concentration can be used to calculate the bulk modulus and shear modulus of the composite as an isotropic polycrystal by averaging over different crystallite orientations. For this purpose, in this work we used the Voigt‒Reuss‒Hill approximation method \cite{Shermergor}:

\begin{eqnarray}\label{eq15}
	k_{VRH} = (k_V + k_R)/2, \nonumber \\
	\mu_{VRH} = (\mu_V + \mu_R)/2,
\end{eqnarray}

\noindent where the subscripts $V$ denote the Voigt averages (upper estimate of the elastic moduli)

\begin{eqnarray}\label{eq16}
	k_V = \dfrac{1}{3} (c_{11} + 2 c_{12}), \nonumber \\
	\mu_V = \dfrac{1}{5}(c_{11} - c_{12} + 3 c_{44}),
\end{eqnarray}

\noindent and the subscripts $R$ denote the Reuss averages (lower estimate)

\begin{eqnarray}\label{eq17}
	k_{R} = \dfrac{1}{3(s_{11} + s_{12})}, \nonumber \\
	\mu_{R} = \dfrac{5}{4(s_{11} - s_{12}) + 3s_{44}}.
\end{eqnarray}

The study showed that when use is made of the values of effective stiffness and compliance constants calculated by the MD method in expressions \eqref{eq16} and \eqref{eq17}, the upper and lower estimates of the elastic properties coincide with sufficient accuracy. The bulk moduli calculated using the Voigt and Reuss approximations coincide with each other up to a rounding error [as should be due to \eqref{eq14}, \eqref{eq16}, and \eqref{eq17}], and the shear moduli differ by no more than 1.4\%.

\section{Results and discussion}

\begin{figure}[ht]
\includegraphics[width=1.0\linewidth]{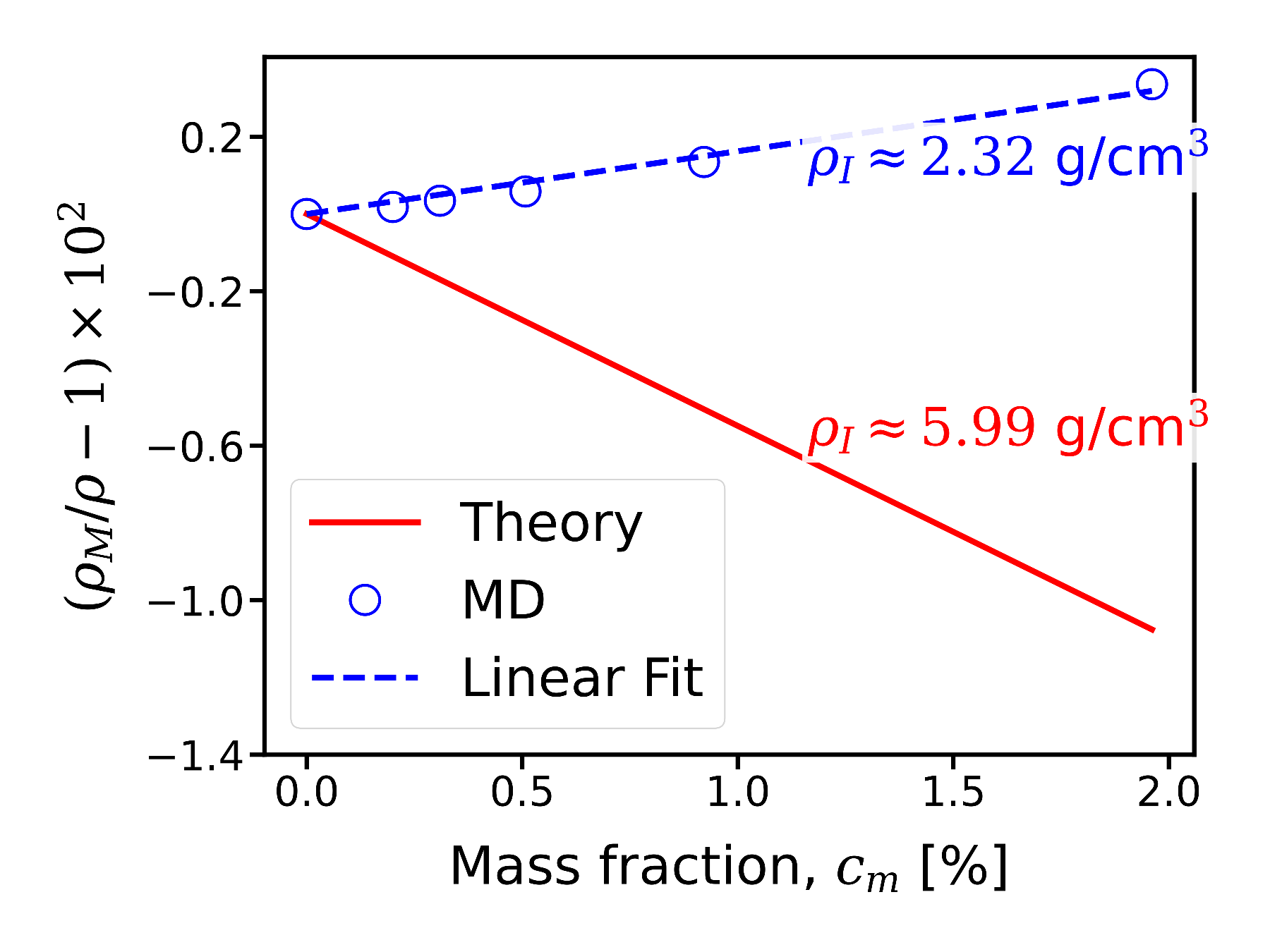}%
\caption{\label{fig2}
Dependence of the inverse effective density of the \alc\ composite on the mass fraction of carbon atoms. The solid line shows the analytical estimate using formula \eqref{eq3} at $R_I = 3.57$~\AA, the circles show the results of MD calculations, and the dashed line shows the approximation of the results of numerical calculations by expression \eqref{eq3}. Next to the lines are the fullerene density values $\rho_I$, the use of which in calculation \eqref{eq3} allows one to obtain the indicated dependence of $\rho(c_m)$.
}%
\end{figure}

Approximation of the $\rho(c_m)$ dependence obtained using the MD method by expression \eqref{eq3} allows us to calculate the unknown parameters of fullerene: density $\rho_I$ and volume $V_I = m_I / \rho_I$. The results of analytical estimates, MD calculations and their approximation are presented in Fig.~\ref{fig2}.

One can see from Fig.~\ref{fig2} that the use of analytical estimates does not make it possible to reproduce the data obtained in MD calculations. According to the results of MD calculations, with increasing concentration of fullerenes, the effective density of the composite decreases, and its reciprocal value, plotted along the ordinate axis in Fig.~\ref{fig2}, increases. The analytical model predicts an increase in density, which is explained by an incorrect calculation of the fullerene volume. Recalculation of the fullerene radius using the results of density approximation obtained in MD calculations yields $R_{I}^{MD} = 4.98$~\AA, which differs markedly from the value of 3.57~\AA\ used in our analytical estimates. To illustrate the reason for this discrepancy, a single-particle distribution function $n(\mathbf{r})$ can be used, constructed for carbon and aluminum atoms in a spherical coordinate system with the origin at the center of mass of the fullerene and averaged over angular coordinates \cite{Balescu}:

\begin{equation}\label{eq18}
	n(r_1) = \dfrac{1}{4 \pi}\int F(\mathbf{r}_1, ..., \mathbf{r}_N) d \mathbf{r}_2 ... d \mathbf{r}_N d \varphi_1 d \theta_1,
\end{equation}

\noindent where $N$ is the total number of particles, $F$ is the distribution function in the configuration space, and $\mathbf{r}_i$ is the radius vector of a particle with number $i$.

\begin{figure}[ht]
\includegraphics[width=1.0\linewidth]{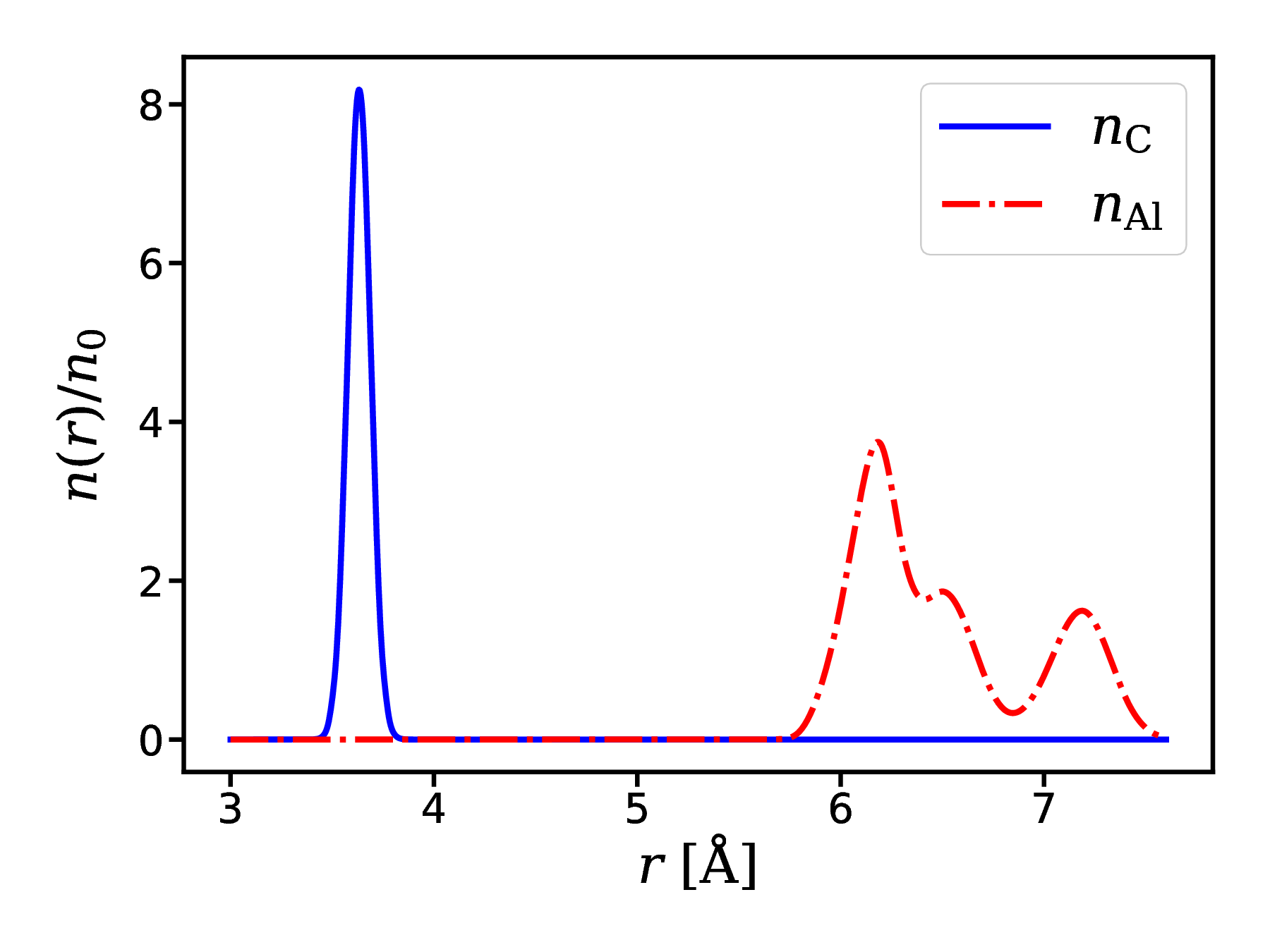}%
\caption{\label{fig3}
Single-particle distribution functions averaged over angular coordinates, plotted for carbon and aluminum atoms relative to the center of mass of the fullerene.
}%
\end{figure}

Figure~\ref{fig3} shows functions $n(r_1)$ plotted for carbon and aluminum atoms and related to the corresponding average densities $n_0 = N/V$. In calculating the density $n_0$ of fullerene, the radius was taken equal to 3.57~\AA, and the volume was calculated using the formula for the volume of a sphere.

Figure~\ref{fig3} shows the positions of the first peaks of the distribution function, approximately corresponding to the average distance from the center of mass to carbon atoms (fullerene radius, $R_I \approx 3.635$~\AA) and to the nearest aluminum atoms (cavity radius, $R_H \approx 6.182$~\AA). It is easy to see that the numerical values are in good agreement with the data from \cite{eletskiui1993fullerenes}, as well as with the results of analytical estimates using formula \eqref{eq12}. At the same time, the value $R_{I}^{MD} = 4.98$~\AA\ obtained by approximating the $\rho(c_m)$ dependence differs little from the average radius $\langle R \rangle = (R_I + R_H)/2 \approx 4.91$~\AA.

The radius of inclusions and mass density of fullerene [and therefore, according to \eqref{eq4} the volume fraction] can be approximately estimated using the analytical model \cite{reshetniak2022effect}: with the fullerene radius $R_I \approx 3.57$~\AA\ and the cavity radius $R_H \approx 5.88$~\AA\ [see formula \eqref{eq12} and explanation for it], we obtain $R_{I}^{MD} \approx \langle R \rangle \approx 4.73$~\AA, $\rho_I = 3 m_I /[4 \pi (R_{I}^{MD})^3] \approx 2.73$~g/cm$^3$. Comparison with the results of approximation of MD calculations demonstrates that the use of the model from paper \cite{reshetniak2022effect} can significantly increase the accuracy of analytical estimates when calculating the dependence of density on the mass fraction of carbon in \alc\ composites. This result can be useful, for example, in analyzing the porosity of nanocomposites.

The values of mass density, components of the stiffness and compliance tensors of the \alc\ composite, calculated by the MD method for various fullerene concentrations, as well as bulk and shear moduli calculated according to \eqref{eq15}‒\eqref{eq17}, are presented in Table.~\ref{tab2}. In addition to the values of mass fractions, the table also shows volume fractions calculated by approximating the $\rho(c_m)$ dependence. Also, for comparison, Table~\ref{tab2} lists experimental data from \cite{frantsevich1982elastic} for pure aluminum at room temperature 300~K.

\begin{table*}[ht] 
\caption{\label{tab2} Mass density of isothermal elastic moduli of aluminum and \alc\ composites at $T = 300$~K as functions of the concentration of fullerenes.}

\begin{ruledtabular}

\begin{tabular}{lcccccccc}
 & $c_m$,~\% & $c$,~\% & $\rho$, g/cm$^3$ & $c_{11}$, GPa & $c_{22}$, GPa & $c_{44}$, GPa & $k$, GPa & $\mu$, GPa \\ \hline
 Experiment~\cite{frantsevich1982elastic} & 0.00 &  0.00 & 2.6889 & 107.3 & 60.8 & 28.3 & 76.3 & 26.2\\
\hline

 \multirow{6}{*}{MD}  & 0.00 & 0.00 & 2.7017 & 103.5 & 55.7 & 22.77 & 71.60 & 23.22 \\
   & 0.20 & 0.22 & 2.7013 & 103.3 & 55.9 & 22.75 & 71.70 & 23.13 \\
   & 0.31 & 0.34 & 2.7008 & 103.1 & 56.1 & 22.74 & 71.75 & 23.06 \\
   & 0.51 & 0.56 & 2.7002 & 103.0 & 56.3 & 22.74 & 71.89 & 22.98 \\
   & 0.93 & 1.06 & 2.6981 & 102.5 & 56.7 & 22.52 & 71.95 & 22.68 \\
   & 2.00 & 2.30 & 2.6927 & 101.6 & 58.0 & 22.38 & 72.53 & 22.14 \\
\end{tabular}

\end{ruledtabular}

\end{table*}

Table~\ref{tab2} shows that for pure aluminum the calculated data are in good agreement with the experimental ones. At the same time, the theoretical value of aluminum density is slightly overestimated compared to the experimental value, and the values of the stiffness constants are underestimated. The model produces the most significant error for the constant $c_{44}$, which in the calculations turns out to be underestimated by 5.6~GPa (about 20\%). The existing discrepancy between theory and experiment may be due to an approximate calculation of the forces of interatomic interactions using the EAM potential. Parameterization of the potential was performed by Mishin et al.~\cite{mishin1999interatomic}, who used the values of the stiffness constants at $T = 0$~K as input data; therefore, zero temperature guarantees a high accuracy in calculating the stiffness constants. According to Mishin et al.~\cite{mishin1999interatomic}, the values of the stiffness constants at $T = 0$~K are $c_{11} = 114$~GPa, $c_{12} = 61.6$~GPa and $c_{44} = 31.6$~GPa, which is exactly consistent with the reference data used in the parameterization~\cite{simmons1971single}. Our calculation at $T = 0$~K yields values of the stiffness constants that coincide with the data from \cite{mishin1999interatomic, simmons1971single} within an error of 5\%. At the same time, since the values of the elastic moduli of aluminum at $T = 300$~K were not included in the training data set for the potential parameterization, the calculation of elastic properties at a given temperature may be associated with an error, as evidenced by the data in Table.~\ref{tab2}.

\begin{figure*}[ht]
\centering
\begin{minipage}{0.49 \linewidth}
	\includegraphics[width=1\linewidth]{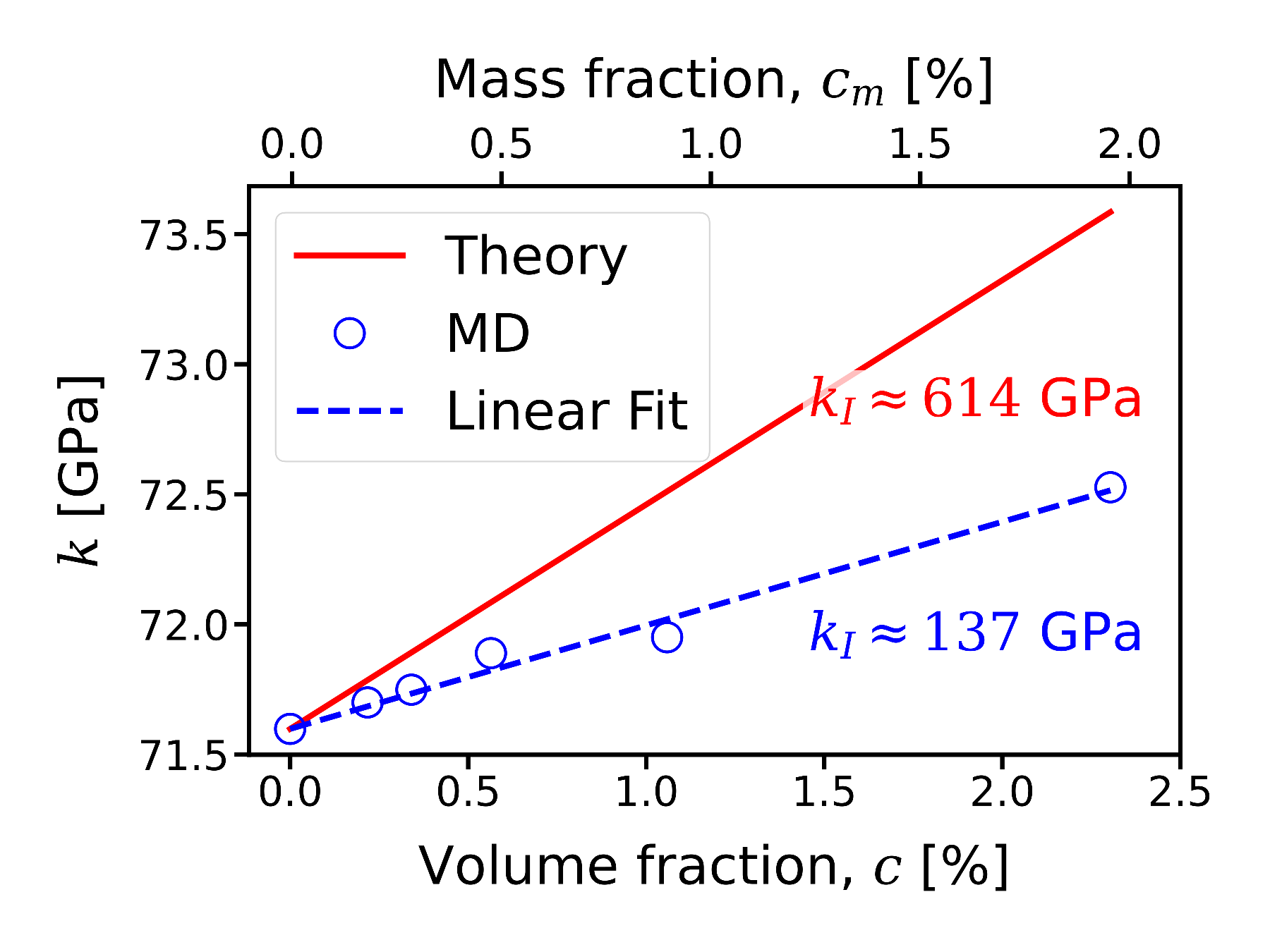}%
\end{minipage}
\begin{minipage}{0.49 \linewidth}
	\includegraphics[width=1\linewidth]{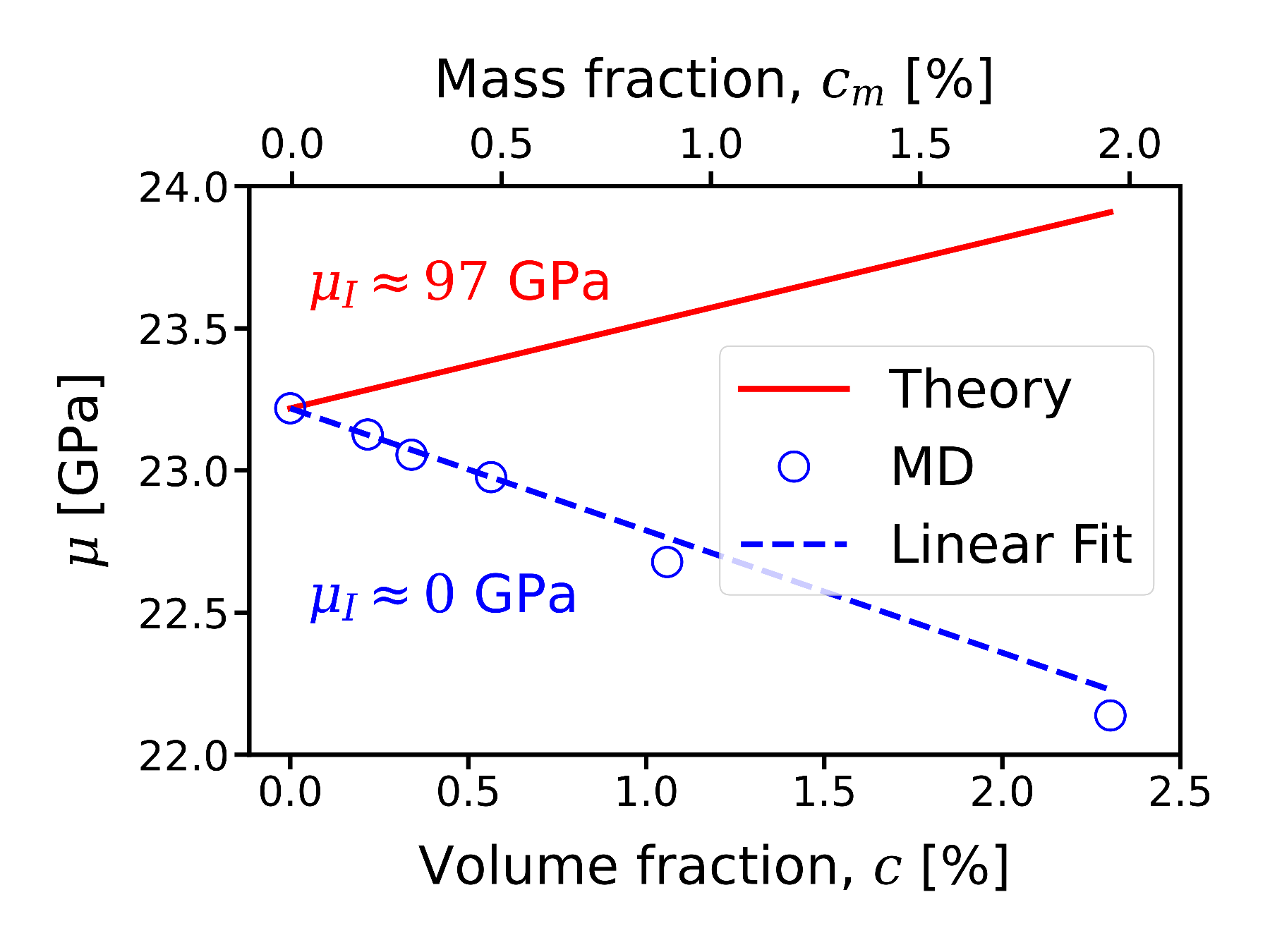}%
\end{minipage}\\
\vspace{0.1cm}

\raggedright \hspace{0.25\linewidth}$(a)$\hspace{0.48\linewidth} $(b)$
\caption{\label{fig4}
Effective moduli of (a) uniform compression and (b) shear of the \alc\ composite as functions of the concentration of fullerenes. Volume concentrations of fullerenes were calculated by approximating the dependence of density on the mass fraction of carbon and correspond to the data from Table~\ref{tab2}. Solid lines indicate the results of estimates using formulas \eqref{eq4} and \eqref{eq5}, in which the elastic moduli are taken from Table~\ref{tab1} for the Tersoff potential (values are indicated in the figure and colored in red). Circles show data obtained by the MD method, and dashed lines show the results of their approximation using formulas \eqref{eq4} and \eqref{eq5}. The values of the elastic moduli of the inclusion obtained by approximating the results of MD calculations by expressions \eqref{eq4} and \eqref{eq5} are colored in blue. 
}%
\end{figure*}

Despite the differences, analysis of the data from Table~\ref{tab2} allows us to determine the effect of fullerene concentration on the elastic properties of the \alc\ composite. Figure~\ref{fig4} shows dependences of the bulk elastic modulus and shear modulus on the concentration of fullerenes.

Figure~\ref{fig4} shows that analytical estimates using formulas \eqref{eq4} and \eqref{eq5} diverge from the results of numerical simulation not only quantitatively, but also qualitatively. It should be noted that the approximations and assumptions used in atomistic MD simulations are more general than those of the theory of heterogeneous media. Continuum calculation uses the division of a deformable object into infinitesimal elements, the elastic properties of which are assumed to correspond to the properties of the given object. This approximation, whose applicability for nano-objects is questionable, is not used in MD simulations.

An attempt to use experimental data \cite{korobov2016mechanical} to test the models, where the velocities of elastic waves were measured for AMg6/C$_{60}$ alloy and then were used to calculate the stiffness constants, does not give satisfactory results. Unfortunately, Korobov et al.~\cite{korobov2016mechanical} measured only the mass fraction of fullerenes, $c_m = 0.3$\%, and did not study the dependence of elastic properties on concentration. In this case, it is possible to compare the results of calculations and the results of work \cite{korobov2016mechanical} for a given concentration. Data from Table~\ref{tab2} indicate that the mass fraction of fullerenes, $c_m = 0.3$\%, in the \alc\ composite approximately corresponds to the volume fraction $c = 0.33$\%. According to Korobov et al.~\cite{korobov2016mechanical}, the values of the stiffness constants for the AMg6/C$_{60}$ composite turned out to be lower than for those for the original alloy. Recalculation of the bulk modulus and shear modulus with formulas \eqref{eq15}‒\eqref{eq17} using data from \cite{korobov2016mechanical} indicates a decrease in the bulk modulus (68.7~GPa) and shear modulus of the composite (25.4~GPa) compared to the parameters of the original alloy (72.5~GPa and 25.9~GPa, respectively). When formulas \eqref{eq5} are used for a given concentration of inclusions, we find that to ensure the specified reduction in the bulk modulus of the composite, the bulk modulus and shear modulus of the inclusion must be negative and equal to ‒25 and ‒15~GPa, respectively. The appearance of negative values of elastic moduli, which have no physical meaning, may indicate porosity of the composite, which cannot be ignored in calculations. At these low concentrations of fullerenes, even a relatively small concentration of pores (about 1\%) can have a more noticeable effect on the elastic modulus of the material than the presence of inclusions. A hydrostatic weighing investigation \cite{korobov2016mechanical} indicates a decrease in the density of the composite compared to the density of the original material by approximately 1\%. Comparison with the data in Table~\ref{tab2} allows us to conclude that this decrease in density is due to the porosity of the composite rather than to the presence of \fullerene\ fullerenes. Estimating the pore concentration with formula \eqref{eq2} yields a value of about 1.3\%, and subsequent calculation of the bulk modulus and shear modulus of the porous material with formulas \eqref{eq5} yields values of 69.6~GPa and 25.3~GPa, which are in good agreement with the measurement results. Thus, taking into account the influence of pores on the effective elastic moduli of the composite makes it possible to describe the obtained experimental data quite well, while estimating the elastic moduli of the inclusion according to the data of~\cite{korobov2016mechanical} turns out to be difficult due to the low concentration of fullerenes, as well as the lack of experimental data on the dependence of the elastic moduli on concentration.

The results obtained in this work indicate a significant decrease in the elastic moduli of the inclusion in the composite compared to analytical estimates and numerical calculations. This effect can be explained by the theory \cite{gurtin1998general,sharma2004size,lurie2006interphase}. The distance between the nearest interacting carbon and aluminum atoms is finite, comparable with the radius of the fullerene, and changes during elastic deformation of the composite. This effect can be observed by analyzing the positions of the peaks of the single-particle distribution function $n(r)$ during all-round compression of the system. Thus, when the system is compressed by 3\%, the position of the maximum of the function $n_{\mathrm{C}} (r)$, which determines the fullerene radius $R_I$, shifts to the left: after compression, $R_I \approx 3.627$~\AA, i.e., approximately 0.22\% less than the original value. The first maximum of the function $n_{\mathrm{Al}} (r)$, the position of which determines the cavity radius $R_H$, shifts by approximately 0.75\% ($R_H \approx 6.136$~\AA) during compression. Consequently, as a result of strain, the boundary atoms of fullerene and aluminum are displaced relative to each other, and on a spherical dividing surface of arbitrary radius $R_I \leq R_S \leq R_H$, the radial component of the strain tensor $\epsilon_{rr}$ experiences a first-order discontinuity. Note that equations \eqref{eq5} were obtained under the assumption that strains at the interface are compatible, and their use does not imply the possibility of relative displacement of the fullerene surfaces and the cavity. The authors of Refs.~\cite{gurtin1998general, sharma2004size, lurie2006interphase} proposed a model of an interphase layer---an elastic medium separating the surfaces of fullerene and aluminum---to take into account the incompatibility of strains. The elastic nature of the strain of the interphase layer is determined by the dependence of the surface energy on the relative displacement of carbon and aluminum atoms. Elastic moduli and the volume of the interphase layer are theoretical parameters, which can be estimated from the results of MD calculations \cite{sharma2004size}. Assuming that the inclusion can be represented as a spherical composite core‒shell particle, where the core is a fullerene, and the shell is an interphase layer, the parameters of which need to be calculated, we calculate the elastic moduli of the interphase layer using the expressions of the theory of elasticity \cite{christensen2012mechanics}:

\begin{equation}\label{eq19}
	k_{IL} = k_L + \dfrac{c_0 (k_{I} - k_L)}{1 + (1-c_0)[(k_{I} - k_L)/(k_L + 4\mu_L /3)]},
\end{equation}

\noindent where $k_{IL}$ is the bulk modulus of the composite particle, $k_I = 614$~GPa is the bulk modulus of fullerene (see Table~\ref{tab1}), $c_0 = R_I^3/R_L^3$ is the volume fraction of fullerene in the composite particle, and the subscript $L$ denotes the interphase layer parameters.

The fact that the shear modulus of a composite particle is zero for the \alc\ system (see Fig.~\ref{fig4}) means that the shear modulus of the interphase layer is also zero. Since the effective bulk modulus of the composite particle, $k_{IL} = k_I^{MD} = 137$~GPa, obtained by approximating the $k(c_m)$ dependence, is significantly less than the bulk modulus of the fullerene calculated using the Tersoff potential and presented in Table~\ref{tab1}, $k_I = 614$~GPa, we can also assume that the value of $k_L/k_I$ is small. Then expression \eqref{eq19} yields

\begin{equation}\label{eq20}
	k_L \approx k_{IL} (1 - c_0).
\end{equation}

After using $R_L = R_{I}^{MD} = 4.98$~\AA\ to calculate $c_0$ (see discussion in Fig.~\ref{fig2}), we obtain $k_L \approx 84$~GPa. If the parameters of interatomic interaction are known, an analytical estimate of the bulk modulus of the interphase layer is possible, which does not require the use of fitting to the known results of MD calculations~\cite{reshetniak2022effect}. In pair interatomic interactions, the adhesion energy of a particle and a matrix is determined from the expression~\cite{Balescu}:

\begin{equation}\label{eq21}
	U = \int n_2 (\mathbf{r}_1, \mathbf{r}_2) u( \left\lvert \mathbf{r}_2 - \mathbf{r}_1 \right\rvert) d \mathbf{r}_1 d \mathbf{r}_2,
\end{equation}

\noindent where $n_2(\mathbf{r}_1, \mathbf{r}_2)$ is the two-particle distribution function, and $\mathbf{r}_1$ and $\mathbf{r}_2$ are the radius vectors of the configuration space points of the first and second phases. An accurate calculation of the adhesion energy of phases, the atoms of which interact through a given pair potential, is possible using the molecular or Monte Carlo method; however, for the analytical assessment of the adhesion energy it is necessary to use simplifying assumptions about the form of the two-particle distribution function. In~\cite{reshetniak2022effect} we used an approximation that neglected the correlated motion of aluminum and carbon atoms. In this case, the two-particle distribution function is written as a product of single-particle functions:

\begin{equation}\label{eq22}
	n_2( \mathbf{r}_1, \mathbf{r}_2) \approx n_{C}(\mathbf{r}_1) n_{Al}(\mathbf{r}_2).
\end{equation}

Single-particle distribution functions for aluminum and carbon were assumed to be isotropic and have the form:

\begin{eqnarray}\label{eq23}
	n_C (r_1) = n_{\mathrm{C}}^{(0)} \delta(r_1 - R), \nonumber \\
	n_{Al} (r_2) = n_{\mathrm{Al}}^{(0)} \theta(r_2 - R_H),
\end{eqnarray}

\noindent where $n_{\mathrm{C}}^{(0)} = 0.3614$~\AA$^{-2}$ is the average surface density of fullerene, $n_{\mathrm{Al}}^{(0)} = 4/a^3 = 0.0602$~\AA$^{-3}$ is the average density of aluminum, $a = 4.05$~\AA\ is the parameter of the face-centered cubic lattice of aluminum, $\delta(r)$ is the Dirac delta function, and $\theta(r)$ is the Heaviside step function. For potential \eqref{eq11}, integral \eqref{eq21} was calculated analytically using approximate expressions \eqref{eq22} and \eqref{eq23}, after which, by differentiating the energy over the volume, we obtain the approximate expression for the bulk modulus of the interphase layer \cite{reshetniak2022effect}:

\begin{equation}\label{eq24}
	k_L = 27 n_{\mathrm{Al}}^{(0)} n_{\mathrm{C}}^{(0)} R \varepsilon \sigma.
\end{equation}

Calculation using formula \eqref{eq24} yields $k_L = 49$~GPa, and the bulk modulus of a composite “core--shell” particle, where the fullerene acts as the core and the elastic interphase layer as the shell, is $k_{IL} = 80$~GPa. Thus, the analytical estimate using formulas \eqref{eq20} and \eqref{eq24}, which does not involve resource-intensive numerical studies, underestimates the value of the effective bulk modulus of the inclusion by approximately 1.7 times, while neglecting the interfacial interaction in calculation of the bulk modulus of fullerene gives a result that is overestimated by 4.5 times. Thus, a comparison of the results of analytical estimates and numerical simulation makes it possible to conclude that the analytical model in the study of elastic deformations in nanocomposites can be used in practice for planning numerical or laboratory experiments, or for a qualitative description of their results. However, the assumptions about the form of the two-particle distribution function do not guarantee high accuracy in calculating the surface energy and bulk modulus of the interphase layer.

\section{Conclusions}

We have studied small deformations of solid-phase inhomogeneous \alc\ systems. Using the molecular dynamics method, we have calculated the dependences of density and elastic moduli on the concentration of fullerenes, and then, analyzed the calculation results within the framework of the theory of heterogeneous media. We have shown that using a heterogeneous medium model for this system is possible, but requires the introduction of an interphase layer between aluminum and carbon phases. The parameters of the interphase layer that are determined by the interaction of Al--C atoms have been calculated by approximating the results of MD calculations. To calculate the effective elastic moduli of the inclusion, we have used a model of a composite core--shell particle, where the fullerene acts as the core and the interphase layer acts as the shell. In using a model of a heterogeneous medium and assuming the boundary between aluminum and fullerene to be ideally thin and rigid, the dependences of the effective parameters of the composite on the concentration of \fullerene\ diverge from the results of numerical MD calculations. In this case, the use of the theory of composites leads to qualitative errors: the theory predicts an increase in the effective moduli of elasticity and density of the composite with an increase in the mass fraction of fullerenes, while the calculation results indicate a monotonic decrease in the functions $\rho (c_m)$ and $\mu (c_m)$.

The Lennard‒Jones potential and a number of approximations that simplify the form of the two-particle distribution function make it possible to analytically calculate the bulk modulus and thickness of the interphase layer for liquid-phase systems within the framework of the model proposed in \cite{reshetniak2022effect}. Simplified analytical consideration of the interphase layer slightly complicates the expressions compared to a two-phase model of a heterogeneous medium, but at the same time allows for a noticeable increase in the accuracy of evaluations. A comparison of analytical estimates with the results of numerical calculations have shown that the use of an analytical model provides high accuracy in calculating the interphase layer thickness (and hence the effective density of the composite), but yields an underestimated value of the bulk modulus. In this work, the error in the analytical estimate of the effective bulk modulus of a composite particle (fullerene and interphase layer) is about 40\%. Note that in the two-phase model, which assumes the absence of a interphase layer and ideal contact between the aluminum and fullerene surfaces, the value of the bulk modulus of inclusions turns out to be overestimated by 4.5 times.

Thus, analytical estimates by the model from \cite{reshetniak2022effect} can be useful in planning laboratory and numerical experiments, as well as in analyzing and explaining the obtained results. The analytical model guarantees high accuracy in calculating the effective density of a nanocomposite, and so its use can be useful for analysis of the material porosity by the hydrostatic weighing method. Increased accuracy in calculating the effective modulus of elasticity of \alc\ composites can be ensured by atomistic simulation methods.

\begin{acknowledgments}
The work was supported by the Ministry of Science and Higher Education of the Russian Federation in the framework of the state assignment in the field of scientific activity (topic FZUN-2024-0004, state assignment of Vladimir State University).
\end{acknowledgments}

\appendix*
\section{Shear modulus of fullerene}

Let the fullerene take the shape of a spheroid as a result of strain, with its volume being unchanged. The semi-axes of the spheroid are related to the fullerene radius by the formulas:

\begin{eqnarray}\label{eqa1}
	a_{p} = R_I(1 - e^2)^{-1/3}, \nonumber \\
	b_{p} = R_I(1 - e^2)^{1/6}
\end{eqnarray}

\noindent for a prolate spheroid (indicated by the subscript $p$), and

\begin{eqnarray}\label{eqa2}
	a_o = R_I(1 - e^2)^{1/3}, \nonumber \\
	b_o = R_I(1 - e^2)^{-1/6}
\end{eqnarray}

\noindent for an oblate spheroid (subscript $o$).

For elastic strains, eccentricity $e \ll 1$. Having chosen a Cartesian coordinate system with the $z$ axis directed along the symmetry axis of the ellipsoid, we write the expression for the strain tensors:

\begin{eqnarray}\label{eqa3}
\epsilon_p = \dfrac{e^2}{6} \left[
\begin{array}{ccc} 
2 & 0 & 0 \\  
0 & -1 & 0 \\ 
0 & 0 & -1 
\end{array}
\right] = \epsilon,
 \nonumber \\
\epsilon_o = \dfrac{e^2}{6} \left[
\begin{array}{ccc} 
-2 & 0 & 0 \\  
0 & 1 & 0 \\ 
0 & 0 & 1 
\end{array}
\right] = -\epsilon.
\end{eqnarray}

The trace of the strain tensor \eqref{eqa3} is equal to zero; therefore, the strain is not related to a change in the volume of the particle and is a shear strain. The surface areas of prolate and oblate spheroids are determined by the formulas:

\begin{eqnarray}\label{eqa4}
	S_p = 2 \pi b_p^2 \left(1 + \dfrac{a_p \arcsin e}{b_p e} \right), \nonumber \\
	S_o = 2 \pi b_o^2 \left(1 + \dfrac{1 - e^2}{e} \arsinh e \right).
\end{eqnarray}

Using the Taylor series expansion, it is easy to show that for the same values of eccentricity, up to terms of the sixth order of smallness, the areas of the oblate and prolate spheroids are equal to each other:

\begin{equation}\label{eqa5}
	S_o = S_p = 4 \pi R_I^2 \left(1 + \dfrac{2}{45} e^4 \right).
\end{equation}

It follows from \eqref{eq6} and \eqref{eqa5} that the shear strain energy has the form

\begin{equation}\label{eqa6}
	W = \dfrac{8}{45} \pi R_I^2 \gamma e^4.
\end{equation}

To calculate the shear modulus, we use the well-known expression for the energy density of the elastic strain of an isotropic body \cite{landau1986theory}:

\begin{equation}\label{eqa7}
	w = \dfrac{W}{V_{I0}} = \dfrac{\lambda}{2} \epsilon_{ll}^2 + \mu \epsilon_{ik}^2,
\end{equation}

\noindent where in the first term on the right side the diagonal components of the strain tensor are summed up, and then the square of the sum is calculated, and in the second term the sum of the squares of all components of the tensor is calculated. Expressions \eqref{eqa3} and \eqref{eqa7} yield

\begin{equation}\label{eqa8}
	w = \mu_I e^4.
\end{equation}

Substituting \eqref{eqa6} into \eqref{eqa8}, and then using \eqref{eq9}, we obtain

\begin{equation}\label{eqa9}
	\mu_I = \dfrac{2 \gamma}{15 R_I} = \dfrac{k_I}{5}.
\end{equation}

\nocite{*}

\bibliography{references}

\end{document}